\documentstyle[12pt]{article}
\begin{document}
\title{Exactly solvable quantum mechanical models with
infinite renormalization of the wave function}

\author{O. Yu. Shvedov
\\
Sub-Dept. of  Quantum Statistics and Field Theory, \\
Dept.  of Physics, Moscow State University, \\
119899, Vorobievy Gory, Moscow, Russia}

\footnotetext{e-mail:shvedov@qs.phys.msu.su}

\maketitle

\begin{flushright}
hep-th/0009036
\end{flushright}

\begin{abstract}
The main  difficulty  of  quantum  field  theory  is  the  problem  of
divergences and renormalization.  However, realistic models of quantum
field  theory are renormalized within the perturbative framework only.
It is important to  investigate  renormalization  beyond  perturbation
theory.  However,  known  models  of  constructive field theory do not
contain such difficulties as  infinite  renormalization  of  the  wave
function.  In  this paper an exactly solvable quantum mechanical model
with such a difficulty is constructed.  This  model  is  a  simplified
analog      of      the     large-$N$     approximation     to     the
$\Phi\varphi^a\varphi^a$-model  in  6-dimensional  space-time.  It  is
necessary  to introduce an indefinite inner product to renormalize the
theory.  The mathematical results of the theory of  Pontriagin  spaces
are  essentially  used.  It  is remarkable that not only the field but
also the canonically conjugated momentum become well-defined operators
after adding counterterms.
\end{abstract}

\def\l#1{\label{#1}}
\def\r#1{(\ref{#1})}
\def\c#1{\cite{#1}}
\def\i#1{\bibitem{#1}}
\def\beq{\begin{equation}}
\def\eeq{\end{equation}}
\def\bea{\begin{eqnarray}}
\def\eea{\end{eqnarray}}
\def\beb#1\l#2\eeb{\begin{equation}
\begin{array}{c} #1  \end{array} \label#2  \end{equation}}
\def\bez{\begin{displaymath}}
\def\eez{\end{displaymath}}
\def\bey#1\eey{\begin{displaymath}
\begin{array}{c}
#1 \end{array}
\end{displaymath}}

\section{Introduction}

{\bf 1.}
An essential  feature  of  realistic  models  of  QFT (such as quantum
electrodynamics, Yang-Mills theory etc.) is the property  of  infinite
renormalization of   the  wave  function.  This  difficulty  leads  to
problems of  canonical  quantization  of   the   theory.   Since   the
coefficient $z$ of the term
$\partial_{\mu} \varphi  \partial_{\mu}  \varphi$  of  the  Lagrangian
diverges, the  momentum  canonically conjugated to the field $\varphi$
should be related with the time derivative of the field $\varphi$ as
\bez
\pi = z\dot{\varphi}.
\eez
If we believe $\varphi$ to be an operator-valued distribution \c{BLT},
its derivative  can  be also interpreted in the same way.  Therefore,
the momentum cannot be viewed even as operator distribution because of
infinite coefficient $z$.

Infinite renormalization  of the wave function is a serious difficulty
in the constructive field theory \c{H,GJ}.  Rigorous  construction  of
mathematical models of QFT have been successful for models with finite
$z$ only.  The $z=\infty$-case leads to serious difficulties (see, for
example, \c{ZS}).

This paper  deals  with  the exactly solvable quantum mechanical model
with infinite renormalization of the wave function.  The Lagrangian of
the model is formally written as
\beq
L =  \frac{z\dot{Q}^2}{2}  -  \frac{m^2Q^2}{2}  +  \sum_{k=1}^{\infty}
\left( \frac{\dot{q}_k^2}{2} - \frac{\Omega_k^2 q_k^2}{2} \right) - gQ
\sum_{k=1}^{\infty} \mu_kq_k.
\l{b1}
\eeq
Here $\mu_k$    are     real     quantities,     while     $\Omega_k$,
$k=\overline{1,\infty}$, is  an  increasing  sequence of real positive
numbers.

Renormalization properties of the model \r{b1} depend on the large-$k$
behavior of the sequence $\mu_k$.

(a) If $\sum_k \mu_k^2/\Omega_k <\infty$,  the model is quantized in a
standard way:  one constructs the Hamiltonian, introduces the creation
and annihilation operators
\bez
q_k = \frac{a_k^++a_k^-}{\sqrt{2\Omega_k}},
\qquad
\dot{q}_k = i \sqrt{\frac{\Omega_k}{2}} (a_k^+-a_k^-)
\eez
with the following commutation relations
\bez
[a_k^{\pm},a_m^{\pm}]=0,\qquad [a_k^-,a_m^+]=\delta_{km}
\eez
and shows  the  obtained  Hamiltonian  to  be  a   correctly   defined
self-adjoint operator.

(b). If $\sum_k \mu_k^2/\Omega_k = \infty$ but
$\sum_k \mu_k^2/\Omega_k^2 < \infty$,  renormalization of vacuum  energy
is necessary.   The   Hamiltonian  is  a  self-adjoint  operator  with
nontrivial domain.

(c) If $\sum_k \mu_k^2/\Omega^2_k = \infty$ but
$\sum_k \mu_k^2/\Omega^3_k  <  \infty$,  it  is  necessary  to perform
renormalization of $m^2$.  The vacuum divergences arise.  They can  be
removed by the Faddeev transformation \c{F}.

(d) If $\sum_k \mu_k^2/\Omega^3_k = \infty$ but
$\sum_k \mu_k^2/\Omega^4_k < \infty$,  there is an
additional difficulty:  the  Stueckelberg  divergences  \c{Stu} arise.
They can be removed by the Faddeev-type transformation \c{MS1}.

(e) If $\sum_k \mu_k^2/\Omega^4_k = \infty$ but
$\sum_k \mu_k^2/\Omega^6_k  <  \infty$,  one  should  perform infinite
renormalization of the wave function $z$.

(f) If $\sum_k \mu_k^2/\Omega^6_k = \infty$,  it is necessary  to  add
new counterterms to the Lagrangian.

In this  paper  the model \r{b1} is mathematically constructed for the
most interesting  case  (e).  The cases (a) - (d) are more trivial and
can be investigated with according to \c{Shv} and \c{Shv1}.

It is interesting that the $z=\infty$ case leads to  indefinite  inner
product in the state space analogously to 
the Lee model \c{B1}, the perturbative Hamiltonian QFT \c{Z}, 
the strongly singular potentials in qunatum mechanics \c{Sr,STS,Sn}.
The state space is the Fock space  associated  with  the  one-particle
Pontriagin space.  The  results  of the general mathematical theory of
Pontriagin spaces \c{P,IK,Bgn,AI} are essentially used.

It will be shown that the expressions
\beq
Q(t), \qquad
zQ(t)- \sum_{k=1}^{\infty}                   \frac{g\mu_k}{\Omega_k^2}
q_k(t)
\l{j1}
\eeq
may be both viewed  as  operator  distributions.  Differentiating  the
second expression,  we  obtain  that  the  momentum $P(t)$ canonically
conjugated to $Q(t)$ becomes an operator distribution after  adding  a
counterterm:
\bez
P(t) - \sum_{k=1}^{\infty} \frac{g\mu_k}{\Omega_k^2}
p_k(t)
\eez

 The  model  of  the  type  \r{b1}  arises  in  the  quantum
probability theory \c{HP,Ch}, in the condensed-matter theory
("polaron model" \c{polar}).
It is also an analog of the model  $\Phi\varphi^a\varphi^a$
of a large number of fields which is viewed in the leading order of
$1/N$-expansion (see, for example, \c{Shv} for more details).

This paper  is  organized   as   follows.   Section   2   deals   with
diagonalization of  the  Hamiltonian.  In section 3 renormalization of
the model is  performed.  Section  4  deals  with  constructing  field
operators and  justifying  the  hypothesis  that   expression   \r{j1}
corresponds to   correctly   defined  operator  distributions  in  the
renormalized theory.

\section{Investigation of the regularized model}

Let us quantize the model \r{b1}. Let $\Lambda$ be a positive integer
regularization parameter.    Perform   a   substitution   $\mu_k   \to
\mu_k^{\Lambda}$, where  $\mu_k^{\Lambda}=\mu_k$  at  $k<\Lambda$  and
$\mu_k^{\Lambda}=0$ at   $k>\Lambda$.   Let  $z$  and  $m^2$  be  also
$\Lambda$-dependent, $z_{\Lambda}$  and  $m_{\Lambda}^2$.   Then   the
Hamiltonian takes the form
\beq
H=  \sum_{mn} \left(
\frac{1}{2} p_m Z^{-1}_{\Lambda,mn} p_n + \frac{1}{2} q_m M^2_{\Lambda,mn}
q_n \right),
\l{b3}
\eeq
where $Z_{\Lambda,mn}$,  $m,n=\overline{0,\infty}$ and $M_{\Lambda,mn}^2$
are matrices of the form
\bez
Z_{\Lambda} =
\left(
\begin{array}{cc}
z_{\Lambda} & 0 \\
0 & 1
\end{array}
\right);
\qquad
M_{\Lambda}^2 =
\left(
\begin{array}{cc}
m_{\Lambda}^2 & g\mu^{\Lambda} \\
g\mu^{\Lambda} & \Omega^2
\end{array}
\right);
\eez
$q_0\equiv Q$,  $p_0 \equiv P$ is a momentum conjugated to $Q$,  $p_k$
are momenta conjugated to $q_k$.

Suppose that            there             exists             operators
$(Z_{\Lambda}^{-1}M_{\Lambda}^2)^{\pm 1/4}$.
After transformation
\beb
q_m = \frac{1}{\sqrt{2}}
\sum_{n=0}^{\infty} ((Z_{\Lambda}^{-1}M_{\Lambda}^2)^{-1/4}
Z_{\Lambda}^{-1})_{mn} (b^+_n+b^-_n),\\
p_m =
\frac{i}{\sqrt{2}}
\sum_{n=0}^{\infty}(M_{\Lambda}^2Z_{\Lambda}^{-1})^{1/4}_{mn}
(b^+_n-b_n^-)
\l{b3*}
\eeb
the Hamiltonian  \r{b3} takes the form
\beq
H_{\Lambda}= \sum_{mn=0}^{\infty}
b_m^+[(Z_{\Lambda}^{-1}M_{\Lambda}^2)^{1/2}Z_{\Lambda}^{-1}]_{mn} b_n^-
\l{b4}
\eeq
up to an additive constant interpreted as vacuum energy which can  be
removed by renormalization.

The canonical commutation relations are written as
\beq
[b_m^{\pm},b_n^{\pm}] = 0, \qquad
[b_m^-,b_n^+] = Z_{\Lambda,mn}.
\l{b5}
\eeq
Choose the Fock representation for the operators
$b_m^{\pm}$. Any state vector can be presented as
\beq
\Psi = \sum_{n=0}^{\infty} \frac{1}{\sqrt{n!}}
\sum_{k_1...k_n} \psi^{(n)}_{k_1...k_n}  b_{k_1}^+  ...   b^+_{k_n}
|0>
\l{b6}
\eeq
where $|0>$ is a vacuum state,  $b_k^-|0>=0$, $\psi^{(n)}_{k_1...k_n}$
are functions of $k_1...k_n$ which are symmetric with respect to their
transpositions. Relations  \r{b5}  imply that the inner product can be
presented as
\beq
(\Phi,\Phi) = \sum_{n=0}^{\infty} \sum_{k_1...k_np_1...p_n}
\psi^{(n)*}_{p_1...p_n} Z_{\Lambda,p_1k_1}...Z_{\Lambda,p_nk_n}
\psi^{(n)}_{k_1...k_n}
\l{b6*}
\eeq
while the Hamiltonian operator acts as
\beq
(H_{\Lambda} \psi)^{(n)}_{k_1...k_n} = \sum_{i=1}^n
\sum_{p_i} (Z_{\Lambda}^{-1}M_{\Lambda}^2)^{1/2}_{k_ip_i}
\psi^{(n)}_{k_1...k_{i-1}p_ik_{i+1} ...k_n}.
\l{b7}
\eeq

Evolution operator can be written as
\bey
(e^{-iH_{\Lambda}t} \psi)^{(n)}_{k_1...k_n} = \\
\sum_{i=1}^n
\sum_{p_1...p_n} (e^{-i(Z_{\lambda}^{-1}M_{\Lambda}^2)^{1/2}t})_{k_1p_1}
... (e^{-i(Z_{\Lambda}^{-1}M_{\Lambda}^2)^{1/2}t})_{k_np_n}
\psi^{(n)}_{p_1...p_n}.
\eey
By ${\cal  P}_{\Lambda}$  we  denote  the  space  of  sets   $\psi_k$,
$k=\overline{0,\infty}$ with the indefinite inner product
\beq
<\psi,\psi>_{\Lambda} = (\psi, Z_{\Lambda} \psi).
\l{c1}
\eeq
We see that the state space is the Fock space associated  with  ${\cal
P}_{\Lambda}$:
\bez
{\cal F}({\cal    P}_{\Lambda})    =    \oplus_{n=0}^{\infty}    {\cal
P}_{\Lambda}^{\vee n},
\eez
where ${\cal P}_{\Lambda}^{\vee n}$ is  the  $n$-th  symmetric  tensor
degree of  the  space  ${\cal  P}_{\Lambda}$  \c{BLT}.  The  evolution
operator is
\bez
e^{-iH_{\Lambda}t} =                             \oplus_{n=0}^{\infty}
(\exp({-i(Z_{\Lambda}^{-1}M_{\Lambda}^2)^{1/2}t}))^{\otimes n}.
\eez

\section{Problem of renormalization}

{\bf 1.} There are several ways to renormalize a quantum field  theory
model. For example, one can first evaluate such vacuum expectations as
Green or Wightman functions  \c{W56,BLT},  $r$-functions  \c{LSZ2}  or
$S$-matrix coefficient functions \c{BS,BMP} for the regularized theory
and consider the limit $\Lambda\to\infty$ for these  quantities.  Then
the Wightman  reconstruction  theorem  \c{W56}  or  its  analog can be
applied.

In the approach based on the dynamical Hamiltonian equations of motion
rather than    $S$-matrix    another    way   to   perform   a   limit
$\Lambda\to\infty$ can be used.  If  $H_{\Lambda}$  is  a  regularized
Hamiltonian acting  in  the  Hilbert space ${\cal H}$,  one can try to
choose such   singular   as   $\Lambda\to\infty$   unitary    operator
$T_{\Lambda}: {\cal   H}   \to{\cal   H}$  ("dressing  transformation"
\c{F,H}) that the operator
\bez
T_{\Lambda}^+ e^{-iH_{\Lambda}t} T_{\Lambda}
\eez
has a strong limit as $\Lambda\to\infty$. The limit
\beq
U(t) =   s-lim_{\Lambda\to\infty}   T_{\Lambda}^+   e^{-iH_{\Lambda}t}
T_{\Lambda}
\l{t1}
\eeq
can be interpreted as  an  evolution  operator  in  the  renormalized
theory.
The difficulty of our case is that different  spaces  ${\cal  F}({\cal
P}_{\Lambda})$ are   considered  at  different  values  of  $\Lambda$.
Another essential feature is that ${\cal F}({\cal  P}_{\Lambda})$  are
not Hilbert spaces but indefinite inner product spaces. Therefore, the
requirement \r{t1} should be modified.

We say that renormalization is performed if:

(i) a  Hilbert  inner  product  is  introduced  on   ${\cal   F}({\cal
P}_{\Lambda})$;

(ii) a  Pontriagin  space  $\cal  L$  ("renormalized  state space") is
specified;

(iii) an  operator  $T_{\Lambda}:  {\cal   L}   \to   {\cal   F}({\cal
P}_{\Lambda})$ is defined;

(iv) for  some  operator  $U(t):{\cal  L} \to {\cal L}$ ("renormalized
evolution operator")      and      any      vector       $\Psi       =
(\psi_0,\psi_1,...,\psi_n,0,0,...)$
\beq
||T_{\Lambda} U(t)  \Psi  -  e^{-iH_{\Lambda}t}   T_{\Lambda}   \Psi||
\to_{\Lambda \to\infty} 0.
\l{t2}
\eeq

Condition \r{t2}  is a modification of condition \r{t1}.  Its physical
meaning is the following.  Suppose that $T_{\Lambda}\Psi$ is chosen to
be an initial state in the regularized theory.  Then state at time $t$
can be approximated by the vector $T_{\Lambda} U(t)\Psi$. The operator
$U(t)$ can be viewed as a renormalized evolution operator.

Note also    that    relation   \r{t2}   means   that   the   operator
$e^{-iH_{\Lambda}t}: {\cal F}({\cal P}_{\Lambda}) \to
{\cal F}({\cal  P}_{\Lambda})$ tends to $U(t):  {\cal L} \to {\cal L}$
in a generalized strong sense \c{Kato, Trotter}.

We will choose ${\cal L} = {\cal F}({\cal P})$,
\bez
T_{\Lambda} = \oplus_{n=0}^{\infty} (P_{\Lambda})^{\otimes n}
\eez
for some Pontriagin space $\cal P$  and  some  operator  $P_{\Lambda}:
{\cal P} \to {\cal P}_{\Lambda}$.

{\bf 2.}  To  introduce  a  Hilbert  inner  product on ${\cal F}({\cal
P}_{\Lambda})$, it  is  sufficient   to   introduce   it   on   ${\cal
P}_{\Lambda}$. The   standard   way   is  the  following  \c{IK}.  Let
$e_{\Lambda}$ be   such   element   of   ${\cal   P}_{\Lambda}$   that
$<e_{\Lambda},e_{\Lambda}>_{\Lambda} <  0$.  Denote by $[e_{\Lambda}]$
the one-dimensional  space  $\{\lambda  e_{\Lambda}  |\lambda\in  {\bf
C}\}$, while  $[e_{\Lambda}]^{\perp}$  is  the  space  of  all vectors
$\psi$ such that $<\psi,e_{\Lambda}>_{\Lambda}  =  0$.  If  the  inner
product is   positively   definite   on  $[e_{\Lambda}]^{\perp}$,  the
indefinite inner product space is of the type $\Pi_1$ \c{P,IK}. We see
that it is true for the case $e_{\Lambda} = (1,0,...)$,  provided that
$z_{\Lambda} <0$ (this condition will be  shown  to  be  satisfied  at
sufficiently large $\Lambda$).  The positive definiteness of the inner
product on $[e_{\Lambda}]^{\perp}$ for arbitrary  $e_{\Lambda}$  is  a
corollary of the general theory of Pontriagin spaces \c{IK}.

The Hilbert inner product is introduced as
\beq
(f,g)_{e_{\Lambda}} =  <f,g>_{\Lambda} +
2 \frac{<f,e_{\Lambda}>_{\Lambda}
<e_{\Lambda},g>_{\Lambda}}{|<e_{\Lambda},e_{\Lambda}>_{\Lambda}|}.
\l{t3}
\eeq
One can notice that $(f,g)_{e_{\Lambda}} =  <f,g>_{\Lambda}$  if  $f,g
\perp e_{\Lambda}$,  $(f,g)_{e_{\Lambda}}  = - <f,g>_{\Lambda}$ if $f,g
\in [e_{\Lambda}]$   and   $(f,g)_{e_{\Lambda}}   =   0$    if    $f\in
[e_{\Lambda}]$, $g\perp   e_{\Lambda}$.   All   topologies  on  ${\cal
P}_{\Lambda}$ that correspond to different  choices  of  $e_{\Lambda}$
are equivalent  \c{IK}.  However,  specification  of  $e_{\Lambda}$ is
important since the convergence requirement \r{t2} is  formulated  in
terms of norms $||\cdot|| \equiv \sqrt{(\cdot,\cdot)_{e_{\Lambda}}}$.

{\bf 3.}  It  seems  to  be physically reasonable to choose the vector
$e_{\Lambda}$ as     an     eigenvector      of      the      operator
$Z_{\Lambda}^{-1}M_{\Lambda}^2$ entering  to  the  Hamiltonian.  Since
$Z_{\Lambda}^{-1}M_{\Lambda}^2$ is a Hermitian operator  with  respect
to the  inner  product  \r{c1},  it  has  according  to the Pontriagin
theorem \c{P}    an    eigenvector     $e_{\Lambda}$     such     that
$<e_{\Lambda},e_{\Lambda}> <0$.   Let   us  find  its  explicit  form.
Equation $Z_{\Lambda}^{-1}      M_{\Lambda}^2      e_{\Lambda}       =
{\varepsilon}_{\Lambda} e_{\Lambda}$ is rewritten as
\bey
m_{\Lambda}^2c_{\Lambda}+g\mu^{\Lambda}_k\phi_{\Lambda,k}
= {\varepsilon}_{\Lambda}z_{\Lambda}c_{\Lambda},\\
g\mu_k^{\Lambda} c_{\Lambda} + \Omega_k^2 \phi_{\Lambda,k}
= {\varepsilon}_{\Lambda}\phi_{\Lambda,k}.
\eey
where $e_{\Lambda} = (c_{\Lambda},\phi_{\Lambda})$.
Therefore, for $\phi_{\Lambda,k}$ one has
\beq
\phi_k = \frac{g\mu_kc}{{\varepsilon}-\Omega_k^2}.
\l{d1}
\eeq
The parameter
${\varepsilon}_{\Lambda}$ obeys the following equation
\beq
{\varepsilon}_{\Lambda} z_{\Lambda}
-          m_{\Lambda}^2          =          g^2           \sum_k
\frac{(\mu_k^{\Lambda})^2}{{\varepsilon}_{\Lambda} -\Omega_k^2}.
\l{d2}
\eeq
For vector \r{d1} $<e_{\Lambda},e_{\Lambda}> < 0$ if and only if
\beq
-b_{\Lambda} \equiv z_{\Lambda} + g^2
\sum_k \frac{(\mu_k^{\Lambda})^2}{({\varepsilon}_{\Lambda}
-\Omega_k^2)^2}<0.
\l{d3}
\eeq
It follows from the Pontriagin theorem  \c{P}  that  eq.\r{d2}  has  a
(real or complex) solution obeying property \r{d3}.

Denote by
\beq
z_{\Lambda,R} = z_{\Lambda} + \sum_k
\frac{(\mu_k^{\Lambda})^2}{\Omega_k^4}; \qquad
m_{\Lambda,R}^2 = m_{\Lambda}^2 -
\sum_k \frac{(\mu_k^{\Lambda})^2}{\Omega_k^2}.
\l{x3}
\eeq
the renormalized values of parameters of the theory.

Eq.\r{d2} can be presented in the following form
\beq
\sum_k \frac{(\mu_k^{\Lambda})^2}{\Omega_k^2}
\left(\frac{1}{{\varepsilon}_{\Lambda} -\Omega_k^2}
+ \frac{1}{\Omega_k^2}\right) = z_{\Lambda,R} - m_{\Lambda,R}^2/
{\varepsilon}_{\Lambda}.
\l{d4}
\eeq
It is  possible  to perform a limit $\Lambda\to\infty$,  provided that
$z_{\Lambda}$ and $m_{\Lambda}^2$ are chosen to make $z_{\Lambda,R}$
and $m_{\Lambda,R}^2$ finite as $\Lambda\to\infty$:
\bez
z_{\Lambda,R} \to z_R, \qquad
m_{\Lambda,R}^2  \to m_{R}^2.
\eez
If $\sum_k \mu_k^2/\Omega_k^4 = \infty$ but
$\sum_k \mu_k^2/\Omega_k^6 < \infty$,  the infinite renormalization of
the wave function is indeed necessary, while $z_{\Lambda}$ is negative
at sufficiently large $\Lambda$.

The following cases should be considered.

(1) $m^2_R>0$.

Eq.\r{d4} has a negative solution ${\varepsilon}<0$ obeying  condition
\r{d3}. Hamiltonian system \r{b3} is unstable.

(2) $m_R^2 <0$, $z_R>0$.

There is  an alternative.  There may be no real solutions of eq.\r{d4}
obeying condition \r{d3}. There may be also 2 real negative solutions.
The smaller one obeys requirement \r{d3}. Hamiltonian system \r{b3} is
also unstable.

(3) $m_R^2\le 0$, $z_R<0$.

Eq. \r{d4} may have no real solutions obeying condition \r{d3} or  may
have a  real  positive  solution  satisfying  requirement \r{d3}.  The
latter case  takes  place  at  sufficiently   small   $|m_R|^2$.   The
Hamiltonian system is stable.

Let us  consider the most interesting latter case only.  Note that the
condition $z_R<0$ arises in investigations of large-$N$ QED \c{PLB}.

If $m_R^2=0$,  the formalism of  the  previous  subsection  should  be
slightly modified                     (the                    operator
$(Z_{\Lambda}^{-1}M_{\Lambda}^2)^{-1/4}$ does  not  exist).  For   the
simplicity, consider the case $m_R^2 \ne 0$ only.

Let us  introduce  more convenient coordinates on the Pontriagin space
${\cal P}_{\Lambda}$ in order to remove divergences from  Hilbert  and
indefinite inner products.  First of all, present any vector $\psi \in
{\cal P}_{\Lambda}$ as
\bez
\psi =
\left( \begin{array}{c} 0 \\ \varphi \end{array} \right)
+ c e_{\Lambda},
\eez
where
\bez
e_{\Lambda} =
\left( \begin{array}{c} 1 \\
\frac{g\mu_k^{\Lambda}}{{\varepsilon}^{\Lambda} - \Omega_k^2}
\end{array} \right)
\eez
One finds:
\bez
(e_{\Lambda}, \psi)      =      -     b_{\Lambda}c
+     \sum_{k=1}^{\infty}
\frac{g\mu_k^{\Lambda} \varphi_k}{{\varepsilon}_{\Lambda} - \Omega_k^2},
\eez
where $b$ has a limit as $\Lambda\to\infty$.  Introduce instead of $c$
new variable $\alpha = - b_{\Lambda}^{-1} (e_{\Lambda},\psi)$:
\bez
\alpha = c- b_{\Lambda}^{-1} \sum_{k=1}^{\infty}
\frac{g\mu_k^{\Lambda}\varphi_k}{{\varepsilon}_{\Lambda} - \Omega_k^2}.
\eez
In terms of new  variables  $(\alpha;  \varphi)$  the  inner  products
\r{c1} and \r{t3} take the form
\beb
<\psi,\psi>_{\Lambda} = -b|\alpha|^2 + <<\varphi,\varphi  >>_{\Lambda}
\\
(\psi,\psi)_{e_{\Lambda}} = b|\alpha|^2 + <<\varphi,\varphi >>_{\Lambda}
\l{t3*}
\eeb
with
\beq
<<\varphi,\varphi>>_{\Lambda} =          (\varphi,\varphi)           +
\frac{1}{b_{\Lambda}}
|\left( \frac{g\mu}{{\varepsilon}-\Omega^2}, \varphi \right)|^2
\l{d5}
\eeq
Formula \r{d5} contains no divergences.

By $\tilde{\cal P}_{\Lambda}$ we denote the Pontriagin space  of  sets
$(\alpha,\varphi)$   with   inner  products  \r{t3*}.  The  introduced
isomorphism   $I_{\Lambda}:   \tilde{\cal   P}_{\Lambda}   \to   {\cal
P}_{\Lambda}$ has the following form:  $I_{\Lambda} : (\alpha,\varphi)
\mapsto (c,\phi)$, where
\beb
c =        \alpha        +       \frac{1}{b_{\Lambda}}
\sum_{k=1}^{\infty}
\frac{g\mu_k^{\Lambda} \varphi_k}{{\varepsilon}_{\Lambda}- \Omega_k^2},\\
\phi_k = c \frac{g\mu_k^{\Lambda}}
{{\varepsilon}_{\Lambda} -\Omega_k^2} + \varphi_k.
\l{t3b}
\eeb

{\bf 5.} It seems to be reasonable to specify the renormalized  states
by sets  $\psi=(\alpha,\varphi)$.  By  $\tilde{\cal  P}$ we denote the
indefinite inner product space of such sets with inner products
\bey
<\psi,\psi> = -b|\alpha|^2 + (\varphi,\varphi)   +    \frac{1}{b}
|\left( \frac{g\mu}{{\varepsilon}-\Omega^2}, \varphi \right)|^2
\\
(\psi,\psi) = b|\alpha|^2 + (\varphi,\varphi)   +    \frac{1}{b}
|\left( \frac{g\mu}{{\varepsilon}-\Omega^2}, \varphi \right)|^2
\eey
with ${\varepsilon}= \lim_{\Lambda\to\infty} {\varepsilon}_{\Lambda}$,
$b= \lim_{\Lambda\to\infty} b_{\Lambda}$.
However, $\tilde{\cal  P}$  cannot be viewed as a state space.  First,
the sequence $\frac{g\mu_k}{{\varepsilon}-\Omega_k^2}$ does not belong
to $l^2$,  so  that one should impose the conditions on $\varphi_k$ at
$k\to\infty$. For example,  one can require $\Omega \varphi \in  l^2$.
Next, the  Euclidean  space  with the inner product $(\cdot,\cdot)$ is
not complete,  so that it is necessary to  consider  the  completeness
$\cal P$ of the space $\tilde{\cal P}$.

Investigate the explicit form of the space $\cal P$ (cf.\c{Sn2}).

Let
$\{(\alpha^{(n)},\varphi^{(n)}) \}$,
$\{(\alpha^{(n)'},\varphi^{(n)'})\}$ be fundamental sequences.
They are  equivalent   \c{KF}   if   $||\psi^{(n)}   -   \psi^{(n)'}||
\to_{n\to\infty} 0$. This means that
\beb
\alpha^{(n)}- \alpha^{(n)'} \to_{n\to\infty} 0; \\
||\varphi^{(n)}-\varphi^{(n)'}|| \to_{n\to\infty} 0; \\
(\frac{g\mu}{\Omega^2}; \varphi^{(n)}-\varphi^{(n)'})
\to_{n\to\infty} 0.
\l{ca2}
\eeb
Furthermore, since the sequence
$\{(\alpha^{(n)},\varphi^{(n)}) \}$ is fundamental, sequences
$\alpha^{(n)}$, $\varphi^{(n)}$ and
$(\frac{g\mu}{\Omega^2}; \varphi^{(n)})$
are also fundamental. Therefore,
\beb
\alpha^{(n)} \to_{n\to\infty} \alpha, \\
\varphi^{(n)} \to_{n\to\infty} \varphi; \\
(\frac{g\mu}{\Omega^2}; \varphi^{(n)}) \to_{n\to\infty} \xi.
\l{ca3}
\eeb
Thus, two fundamental sequences are equivalent if and only if
$\alpha'=\alpha$,  $\varphi'=\varphi$, $\xi'=\xi$.

Let us show now that for any set
$(\alpha,\xi,\varphi)$ there exists  a  fundamental  sequence  obeying
conditions \r{ca3}.   Note  that  any  sequence  obeying  requirements
\r{ca3} is fundamental. It is sufficient to consider two partial cases:

(i0  $\varphi=0$;

(ii) $ \alpha=\xi=0$.

Denote $\chi_k= g\mu_k/({\varepsilon}-\Omega_k^2)$.  For the case (i),
set
\bey
\alpha^{(n)}=\alpha,\\
\varphi^{(n)}_k = \xi \chi_k/\sum_{k=0}^n |\chi_k|^2, \qquad k\le n.\\
\varphi^{(n)}_k = 0, \qquad k>n.
\eey
For the  case  (ii),  it  is  sufficient  to check that the set of all
vectors $\varphi \in l^2$ satisfying the relations
\beq
(\chi,\varphi)=0,
\l{ca4}
\eeq
is dense in $l^2$.  To prove this property, it is sufficient to notice
that any finite vector $\varphi$ can be approximated by a sequence
$\varphi^{(n)} \to \varphi$ obeying requirement \r{ca4}:
\bez
\varphi_k^{(n)} =
\left\{
\begin{array}{c}
\varphi_k -    \frac{(\chi,\varphi)}{\sum_{k=0}^n   |\chi_k|^n}\chi_k,
\qquad k\le n;\\
\varphi_k, \qquad k>n
\end{array}
\right.
\eez
Thus, the renormalized state space $\cal P$ is a space of sets
$(\alpha,\varphi,  \xi)$,  where  $\varphi\in  l^2$,
$\alpha\in {\bf C}$,  $\xi\in{\bf C}$.

The following inner products are introduced in ${\cal P}$:
\beb
(\psi,\psi)= b|\alpha|^2 + (\varphi,\varphi) + {b}^{-1} |\xi|^2 \\
<\psi,\psi>= - b|\alpha|^2 + (\varphi,\varphi) + {b}^{-1} |\xi|^2
\l{t3m}
\eeb

{\bf 6.}  Let  us construct the mapping $\tilde{P}_{\Lambda} :{\cal P}
\to \tilde{\cal P}_{\Lambda}$ which transforms the renormalized  state
$(\alpha,\xi,\varphi)$ to         the         regularized        state
$(\alpha_{\Lambda},\varphi_{\Lambda})$. Choose it in such a way that
\beb
\alpha_{\Lambda} \to_{\Lambda\to\infty} \alpha,
\qquad
||\varphi_{\Lambda} - \varphi|| \to_{\Lambda\to\infty} 0, \\
\left( \frac{g\mu^{\Lambda}}{{\varepsilon}^{\Lambda}    -   \Omega^2},
\varphi_{\Lambda}  \right) \to_{\Lambda\to\infty} \xi.
\l{t4}
\eeb
The mapping $P_{\Lambda}:  {\cal P} \to {\cal P}_{\Lambda}$ will  have
the form $P_{\Lambda} = I_{\Lambda} \tilde{P}_{\Lambda}$.

The following proposition is a direct corollary of eqs.\r{t3*}.

{\bf Proposition    1.}    {\it     Let     $\tilde{P}_{\Lambda}     :
(\alpha,\xi,\varphi) \mapsto (\alpha_{\Lambda},\varphi_{\Lambda})$ be a
mapping satisfying requirements \r{t4}. Then
\bez
||(\tilde{\alpha}_{\Lambda},\tilde{\varphi}_{\Lambda}) -
\tilde{P}_{\Lambda}(\alpha,\xi,\varphi) || \to_{\Lambda\to\infty} 0
\eez
if and  only if $(\tilde{\alpha}_{\Lambda},\tilde{\varphi}_{\Lambda})$
obeys requirements \r{t4}.
}

Proposition 1 tells us that the form of operator $T_{\Lambda}$ obeying
requirements \r{t4} is not important.

By $Q_{\Lambda} : \tilde{\cal P}_{\Lambda} \to {\cal P}$ we denote the
operator $Q_{\Lambda} : (\alpha_{\Lambda}, \varphi_{\Lambda} ) \mapsto
(\alpha_{\Lambda}',\xi_{\Lambda}',\varphi_{\Lambda}')$ of the form
\bez
\alpha_{\Lambda}' = \alpha_{\Lambda},
\qquad
\varphi_{\Lambda}' = \varphi_{\Lambda}, \qquad
\xi_{\Lambda}' =
\left( \frac{g\mu^{\Lambda}}{{\varepsilon}^{\Lambda}    -   \Omega^2},
\varphi_{\Lambda}  \right).
\eez
Proposition 1 can be reformulated as

{\bf Proposition  1'}.  {\it
Let  $(\alpha_{\Lambda},\varphi_{\Lambda})  \in
\tilde{\cal P}_{\Lambda}$. Then
$||(\alpha_{\Lambda},\varphi_{\Lambda})|| \to 0 $ if and only if
$Q_{\Lambda}                       (\alpha_{\Lambda},\varphi_{\Lambda})
\to_{\Lambda\to\infty} 0$.}

Eq.\r{t4} also implies

{\bf Proposition 2.}
\beq
s-lim_{\Lambda\to\infty} Q_{\Lambda} \tilde{P}_{\Lambda} = 1.
\l{t4a}
\eeq

We will also require that
\beq
<\tilde{P}_{\Lambda} \psi,      \tilde{P}_{\Lambda}      \tilde{\psi}>
\to_{\Lambda\to\infty} <\psi,\tilde{\psi}>, \qquad
||\tilde{P}_{\Lambda}|| \le A = const
\l{t4n}
\eeq
for some $\Lambda$-independent $a$.

The explicit form of the mapping $\tilde{P}_{\Lambda} :  (\alpha, \xi,
\varphi) \mapsto (\alpha_{\Lambda},  \varphi_{\Lambda})$ can be chosen
as
\bez
\varphi_{\Lambda} =    \varphi    +    \frac{\chi^{\Lambda}   (\xi   -
(\chi^{\Lambda}, \varphi))}{(\chi^{\Lambda},\chi^{\Lambda})}, \qquad
\alpha_{\Lambda} = \alpha,
\eez
where $\chi^{\Lambda}  = \frac{g\mu^{\Lambda}}{{\varepsilon}^{\Lambda}
- \Omega^2}$.   The   properties   $\alpha_{\Lambda}   \to    \alpha$,
$(\chi^{\Lambda},\varphi^{\Lambda}) =   \xi   \to   \xi$  at  $\Lambda
\to\infty$ are evident. Since $||\chi^{\Lambda}|| \to \infty$, one has
\bez
s-lim_{\Lambda\to\infty}        \frac{\chi^{\Lambda}}{(\chi^{\Lambda},
\chi^{\Lambda})} =  0.
\eez
It is sufficient then to check that
\beq
lim_{\Lambda\to\infty}
\frac{(\chi^{\Lambda},\varphi)}{||\chi^{\Lambda}||}, \qquad    \varphi
\in l^2.
\l{t5}
\eeq
For finite sequences $\varphi$, property \r{t5} is evident. It follows
from the standard theorems of functional analysis \c{KA} that property
\r{t5} is then satisfied for all $\varphi \in l^2$.

{\bf 7.} To check property \r{t2}, it is convenient to investigate the
resolvent of  the  operator $Z_{\Lambda}^{-1} M_{\Lambda}^2$.  To find
its explicit form,
\bez
(\lambda + Z_{\Lambda}^{-1}M_{\Lambda}^2)^{-1}:
\left( \begin{array}{c} c \\ \phi \end{array} \right)
\mapsto
\left( \begin{array}{c} c \\ \phi \end{array} \right)
\eez
one should solve the system of equations
\bey
(\lambda z_{\Lambda}  +  m_{\Lambda}^2)c'  +   g\mu^{\Lambda}\phi'   =
z_{\Lambda}c;\\
g\mu^{\Lambda} c' + (\lambda + \Omega^2) \phi ' = \phi.
\eey
Therefore,
\bey
c' =       a_{\Lambda}[z_{\Lambda}c       -       g\sum_{k=1}^{\infty}
\frac{\mu^{\Lambda}_k\phi_k}{\lambda  +
\Omega_k^2}],\\
\phi '  =  \frac{1}{\lambda  +  \Omega^2} \phi - \frac{g\mu^{\Lambda}}
{\lambda +
\Omega^2} c'
\eey
with
\bez
a_{\lambda} =   (\lambda   z_{\Lambda}   +   m_{\Lambda}^2
-   \sum_k   \frac{g^2(\mu_k^{\Lambda})^2}{\lambda  +
\Omega_k^2})^{-1}.
\eez
Making use of the $(\alpha,\varphi)$-coordinates, one obtains
\bey
\alpha'=\frac{1}{\lambda + {\varepsilon}_{\Lambda}} \alpha;\\
\varphi' =      \frac{1}{\lambda     +     \Omega^2}     \varphi     +
\frac{g\mu^{\Lambda}
({\varepsilon}_{\Lambda}+\lambda)a_{\Lambda} }
{({\varepsilon}_{\Lambda}-\Omega^2)  (\lambda
+ \Omega^2)} \left(
\frac{g\mu^{\Lambda}}{\lambda + \Omega^2}; \varphi \right)
\eey
(eq.\r{d2} is taken into account).  Thus,  the explicit  form  of  the
operator $I_{\Lambda}^{-1}       (\lambda      +      Z_{\Lambda}^{-1}
M_{\Lambda}^2)^{-1} I_{\Lambda}     :     (\alpha,\varphi)     \mapsto
(\alpha',\varphi')$ is found.
This  operator  is  Hermitian  at  real values of
$\lambda$ with  respect  to  Hilbert  and  indefinite  inner  products
\r{t3*}.

{\bf 8.}   Investigate   now   the   behavior  of  the  resolvent  at
$\Lambda\to\infty$. Consider the operator $Q_{\Lambda}I_{\Lambda}^{-1}
(\lambda +   Z_{\Lambda}^{-1}   M_{\Lambda}^2)^{-1}   I_{\Lambda}    :
(\alpha,\varphi) \mapsto   (\alpha',\xi',\varphi')$   which   can   be
presented as
\beb
\alpha'=\alpha,\\
\varphi' = \frac{1}{\lambda+ \Omega^2} \varphi - \frac{g\mu^{\Lambda}
({\varepsilon}_{\Lambda} + \lambda)a_{\Lambda}}
{({\varepsilon}_{\Lambda} -\Omega^2)(\lambda          +
\Omega^2)}\xi \\ +      \frac{g\mu^{\Lambda}      ({\varepsilon}_{\Lambda}
+\lambda)^2      a_{\Lambda}}
{({\varepsilon}_{\Lambda}-\Omega^2)(\lambda + \Omega^2)} \left(
\frac{g\mu^{\Lambda}}
{({\varepsilon}_{\Lambda} - \Omega^2) (\lambda + \Omega^2)}, \varphi
\right), \\
\xi' =
- (\lambda+{\varepsilon}_{\Lambda}) a_{\Lambda}b_{\Lambda} \left(
\frac{g\mu^{\Lambda}}
{({\varepsilon}_{\Lambda} - \Omega^2) (\lambda + \Omega^2)}, \varphi
\right) \\ - \sum_{k=1}^{\infty} \frac{g^2(\mu^{\Lambda}_k)^2
({\varepsilon}_{\Lambda}+\lambda)a_{\Lambda}}
{({\varepsilon}_{\Lambda} - \Omega^2)^2 (\lambda+ \Omega^2)} \xi
\l{t6}
\eeb
(eq.\r{d2} is used), where $\xi= (\frac{g\mu^{\Lambda}}
{{\varepsilon}_{\Lambda} - \Omega^2},
\varphi)$.

Denote the   mapping   \r{t6}   as    $(\lambda+H_{\Lambda})^{-1}    :
(\alpha,\xi,\varphi) \mapsto  (\alpha',  \xi'  ,  \varphi')$.  It is a
resolvent of a positive self-adjoint operator.  We have  obtained  the
following proposition.

{\bf Proposition 3.} {\it The following relation is satisfied:
\bez
Q_{\Lambda} I_{\Lambda}^{-1}     (\lambda      +      Z_{\Lambda}^{-1}
M_{\Lambda}^2)^{-1} I_{\Lambda}   =   (\lambda   +   H_{\Lambda})^{-1}
Q_{\Lambda}.
\eez
}

We see  that  the operator $(\lambda + H_{\Lambda})^{-1}$ has a strong
limit $(\lambda  +  H)^{-1}$  being  a   resolvent   of   a   positive
self-adjoint operator. General results of \c{Faris,Kato} tell us that
the following statement is satisfied.

{\bf Proposition 4.} {\it Let $f$ be a bounded Borel function. Then
\bez
s-lim_{\Lambda\to\infty} f(H_{\Lambda}) = f(H).
\eez
}

Proposition 3 also implies that

{\bf Proposition 5.} {\it For any bounded Borel function $f:{\bf R}\to
{\bf R}$
\bez
Q_{\Lambda} I_{\Lambda}^{-1}     f(Z_{\Lambda}^{-1}     M_{\Lambda}^2)
I_{\Lambda} = f(H_{\Lambda}) Q_{\Lambda}.
\eez
}

Let us prove relation \r{t2}.

{\bf Proposition  6.} {\it For any bounded Borel function $f:  {\bf R}
\to {\bf R}$ and any $\psi \in {\cal P}$
\bez
||f(Z_{\Lambda}^{-1} M_{\Lambda}^2)  P_{\Lambda}  \psi  -  P_{\Lambda}
f(H) \psi || \to_{\Lambda\to \infty} 0.
\l{t7}
\eez
}

{\bf Proof.} Since $I_{\Lambda}:  \tilde{\cal P}_{\Lambda}  \to  {\cal
P}_{\Lambda}$ is an isomorphism,  proposition 1' implies that relation
\r{t7} is satisfied if and only if
\bez
||Q_{\Lambda} I_{\Lambda}^{-1}
f(Z_{\Lambda}^{-1} M_{\Lambda}^2) I_{\Lambda}
\tilde{P}_{\Lambda}  \psi  -  Q_{\Lambda} \tilde{P}_{\Lambda}
f(H) \psi || \to_{\Lambda\to \infty} 0.
\eez
It follows from proposition 5 that this relation can be rewritten as
\beq
|| f(H_{\Lambda}) Q_{\Lambda}
\tilde{P}_{\Lambda}  \psi  -  Q_{\Lambda} \tilde{P}_{\Lambda}
f(H) \psi || \to_{\Lambda\to \infty} 0.
\l{t8}
\eeq
It follows from propositions 2 and  4  that  $s-lim_{\Lambda\to\infty}
f(H_{\Lambda})Q_{\Lambda} \tilde{P}_{\Lambda} = f(H)$, i.e.
\beq
|| f(H_{\Lambda}) Q_{\Lambda}
\tilde{P}_{\Lambda}  \psi  -
f(H) \psi || \to_{\Lambda\to \infty} 0.
\l{t9}
\eeq
Proposition 2 also implies that
\beq
|| f(H)  \psi  -  Q_{\Lambda} \tilde{P}_{\Lambda}
f(H) \psi || \to_{\Lambda\to \infty} 0.
\l{t10}
\eeq
Combining eqs.\r{t9}  and  \r{t10},   we   obtain   relation   \r{t8}.
Proposition 6 is proved.

{\bf 9.} Thus, we have constructed the renormalized state space
${\cal L} = {\cal
F}({\cal P})$.  The "one-particle" renormalized space is chosen to be a
space of  sets $(\alpha,\xi,\varphi)$ with the inner products \r{t3m},
the Hamiltonian operator is also defined by specifying  the  resolvent
$(\lambda+H)^{-1}$. The evolution operator $U(t)$ entering to eq.\r{t2}
has the form
$$
U(t) = \oplus_{n=0}^{\infty} (e^{-iH^{1/2}t})^{\otimes n}.
$$

\section{"Field" operators}

Let us construct now Heisenberg field operators $Q(t)$,  $q_k(t)$  and
their linear  combinations  in  the renormalized theory.  According to
eqs.\r{b3*}, they should be expressed via  creation  and  annihilation
operators. Let us remind their definition (see, for example, \c{BLT}).

The set  of  all  vectors  $\psi^{\otimes n}$ is a total set in ${\cal
P}^{\vee n}$.  By
$b_n^-(\gamma): {\cal P}^{\vee n} \to {\cal P}^{\vee (n-1)}$,
$b_n^+(\gamma): {\cal   P}^{\vee   (n-1)}   \to  {\cal  P}^{\vee  n}$,
${\gamma}
\in {\cal  P}$  we  denote  the  linear  operators  which are uniquely
defined from the relations
\bey
b_n^-({\gamma}) \psi^{\otimes  n}  = \sqrt{n} ({\gamma},\psi) \psi^{\otimes
(n-1)},\\
b_n^+(\gamma) \psi^{\otimes   (n-1)}   =   n^{-1/2}   \sum_{j=0}^{n-1}
\psi^{\otimes j} \otimes \gamma \otimes \psi^{\otimes (n-1-j)}.
\eey
Moreover,
$||b_n^{\pm}(\gamma)|| \le n^{1/2} ||\gamma||$.
By $b^{\pm}(\gamma):  {\cal  F}({\cal  P})  \to {\cal F}({\cal P})$ we
denote the operators
\bez
(b^+(\gamma)\Psi)_n = b_n^+(\gamma) \Psi_{n-1},
\qquad
b^-(\gamma)\Psi)_{n-1} = b_n^+(\gamma) \Psi_{n}
\eez
which are defined on the set of all finite vectors of the Fock space.

{\bf Proposition 7.} {\it The following relations are satisfied:
\bez
b^+(P_{\Lambda}\gamma) T_{\Lambda} = T_{\Lambda} b^+(\gamma),
\qquad
b^-(\gamma_{\Lambda}) T_{\Lambda}            =             T_{\Lambda}
b^-(P_{\Lambda}^+\gamma_{\Lambda}),
\eez
for arbitrary  $\gamma_{\Lambda} \in {\cal P}_{\Lambda}$,  $\gamma \in
{\cal P}$.
}

The proof is straightforward.

{\bf Proposition    8.}   {\it   Let   $\gamma_{\Lambda}   \in   {\cal
P}_{\Lambda}$, $\gamma \in {\cal P}$, $\Psi$ is a finite vector of the
renormalized Fock space ${\cal F}({\cal P})$ and
\beq
||\gamma_{\Lambda} - P_{\Lambda} \gamma|| \to_{\Lambda \to\infty} 0.
\l{m1}
\eeq
Then
\beq
||b^{\pm}(\gamma_{\Lambda})T_{\Lambda} - T_{\Lambda}  b^{\pm}(\gamma))
\Psi || \to_{\Lambda\to\infty} 0.
\l{m2}
\eeq
}

{\bf Proof.} For the creation operator, eq.\r{m2} means that
\bez
||b^+(\gamma_{\Lambda} - P_{\Lambda}\gamma) \Psi||
\to_{\Lambda\to\infty} 0.
\eez
Let $\Psi = (\psi_0,\psi_1,...,\psi_n,0,0,...)$. Then
\bez
||b^+(\gamma_{\Lambda} -     P_{\Lambda}\gamma)     \Psi     ||    \le
max(1,||P_{\Lambda}||^n) ||\Psi||      n^{1/2}      ||\gamma_{\Lambda}
-P_{\Lambda} \gamma|| \to_{\Lambda\to\infty} 0.
\eez
For the annihilation operator, it is necessary to check that
\beq
||T_{\Lambda} b^-(P_{\Lambda}^+\gamma_{\Lambda}   -  \gamma)  \Psi  ||
\to_{\Lambda\to\infty} 0
\l{m3}
\eeq
for $\Psi \in {\cal P}^{\vee n}$.  The Banach-Steinhaus theorem \c{KA}
implies that it is sufficient to prove property  \r{m3}  for  $\Psi  =
\psi^{\otimes n}$,  $\psi  \in {\cal P}$.  This property is correct if
and only if
\bez
(P_{\Lambda}^+\gamma_{\Lambda} - \gamma,  \psi) \to_{\Lambda\to\infty}
0,
\eez
i.e. $(\gamma_{\Lambda},P_{\Lambda}\psi)        \to_{\Lambda\to\infty}
(\gamma,\psi)$. Eqs.\r{m1}  and  \r{t4n}   confirms   this   property.
Proposition is proved.

Propositions 1' and 8 imply the following corollary.

{\bf Proposition    9.}   {\it   Let   $\gamma_{\Lambda}   \in   {\cal
P}_{\Lambda}$, $\gamma \in {\cal P}$, $\Psi$ is a finite vector of the
renormalized Fock space ${\cal F}({\cal P})$ and
\beq
||Q_{\Lambda} I_{\Lambda}^{-1}\gamma_{\Lambda} -
 \gamma|| \to_{\Lambda \to\infty} 0.
\l{m4}
\eeq
Then property \r{m2} is satisfied.
}

We see   that   the   operator   $b^{\pm}(\gamma_{\Lambda})$   in  the
regularized theory corresponds to the  operator  $b^{\pm}(\gamma)$  in
the renormalized  theory.  One  can say $b^{\pm}(\gamma_{\Lambda}) \to
b^{\pm}(\gamma)$ in generalized strong sense \c{Kato, Trotter}.

Note that the linear combinations $\sum_k b_k^+\zeta_k$ and
$\sum_k b_k^-\zeta^*_k$  of  the  operators $b_k^{\pm}$ \r{b3*} can be
presented as  $b^+(\zeta)$  and  $b^-(\zeta)$   correspondingly   with
$\zeta \in {\cal P}_{\Lambda}$.

Consider the   linear    combination
$\sum_{k=0}^{\infty}    q_k(t) \chi_k^{\Lambda}$
in   the   regularized   theory.   It  follows  from
eqs.\r{b3*} that
\bez
\sum_{k=0}^{\infty} q_k(t)  \chi_k^{\Lambda} = b^+(\gamma_{\Lambda}^t)
+ b^-(\gamma_{\Lambda}^t)
\eez
with
\bez
\gamma_{\Lambda}^t =                                \frac{1}{\sqrt{2}}
e^{i(Z_{\Lambda}^{-1}M_{\Lambda}^2)^{1/2}t}
(Z_{\Lambda}^{-1}M_{\Lambda}^2)^{-1/4} Z_{\Lambda}^{-1} \chi_{\Lambda}.
\eez
Propositions 4,5 and 9 imply the following statement.

{\bf Proposition 10.} {\it Let
\beq
Q_{\Lambda}I_{\Lambda}^{-1}Z_{\Lambda}^{-1}             \chi_{\Lambda}
\to_{\Lambda\to\infty} \gamma,
\l{j2}
\eeq
$\Psi$ is a finite vector of the
renormalized Fock space ${\cal F}({\cal P})$. Then
\bez
||(\sum_{k=0}^{\infty} q_k(t)     \chi_k^{\Lambda}    T_{\Lambda}    -
T_{\Lambda} q_t(\gamma)) \Psi|| \to_{\Lambda\to\infty} 0
\eez
with
\bez
q_t(\gamma) = b^+(\frac{1}{\sqrt{2}} e^{iH^{1/2}t} H^{-1/4} \gamma)
+ b^-(\frac{1}{\sqrt{2}} e^{iH^{1/2}t} H^{-1/4} \gamma)
\eez
}

Let us   write  down  the  explicit  form  of  condition  \r{j2}.  Let
$\chi_{\Lambda} = \left(\begin{array}{c} c_{\Lambda} \\ \phi_{\Lambda}
\end{array} \right)$, $\gamma= (\alpha,\xi,\varphi)$,
\bez
\varphi_{\Lambda} =   \phi_{\Lambda}  -  z_{\Lambda}^{-1}  c_{\Lambda}
\frac{g\mu^{\Lambda}}{{\varepsilon}_{\Lambda} - \Omega^2}.
\eez
Eq.\r{j2} means that
\beb
||\varphi_{\Lambda} -\varphi|| \to_{\Lambda\to 0} 0,\\
(\frac{g\mu^{\Lambda}}{{\varepsilon}_{\Lambda}- \Omega^2})
\to_{\Lambda\to 0} \xi, \\
z_{\Lambda}^{-1} c_{\Lambda}
\to_{\Lambda\to 0} \alpha + \xi/b.
\l{j3}
\eeb
We see     that     the     operator
$\sum_{k=0}^{\infty}    q_k(t)
\gamma_k^{\Lambda}$
in  the  regularized  theory  corresponds  to  the
operator $q(\gamma)$   in   the  renormalized  theory,  provided  that
requirements \r{j3} are satisfied.

{\bf Example 1.} Let $\Omega \phi \in l^2$,  $\Omega\phi_{\Lambda} \to
\Omega \phi$, $c_{\Lambda} = 0$.  Then  the  expression
$\sum_{k=1}^{\infty} q_k(t)  \phi_k$  corresponds  to   the   operator
$q(\alpha,\xi,\phi)$ in the renormalized theory, where
\bez
\alpha= - \frac{\xi}{b}, \qquad \xi = \left(
\frac{g\mu}{\Omega({\varepsilon}-\Omega^2)}, \Omega\phi
\right), \qquad \varphi = \phi.
\eez

{\bf Example 2.} Let  $c_{\Lambda}=1$,  $\phi_{\Lambda}=0$.  Then  the
expression
$\sum_{k=0}^{\infty}    q_k(t)
\gamma_k^{\Lambda}$
takes the form $Q(t)$. For this case $\alpha+ \xi/\beta=0$,
\bez
\varphi_{\Lambda} =                   -                    z_{\Lambda}
\frac{g\mu^{\Lambda}}{{\varepsilon}_{\Lambda} - \Omega^2},
\eez
so that $||\varphi_{\Lambda}|| \to_{\Lambda\to\infty} 0$,
\bez
(\frac{g\mu^{\Lambda}}{{\varepsilon}_{\Lambda} -
\Omega^2}, \varphi_{\Lambda})   =   -z_{\Lambda}^{-1}   (-b_{\Lambda}-
z_{\Lambda}) \to_{\Lambda\to\infty} 1
\eez
Thus, $\xi=1$, so that $\alpha= -1/\xi$. We see that
\bez
Q(t) \to q_t(-1/b,1,0)
\eez

{\bf Example 3.} Let  $c_{\Lambda}  =z_{\Lambda}$,  $\phi_{\Lambda}  =
\frac{g\mu^{\Lambda}}{{\varepsilon}_{\Lambda} -    \Omega^2}$.    Then
$\varphi_{\Lambda} =  0$,   $\xi=0$,   $\alpha+\xi/b=1$.   Thus,   the
expression
\bez
z_{\Lambda} Q(t)    +    \sum_{k=1}^{\infty}    \frac{g\mu^{\Lambda}_k
q_k(t)}{{\varepsilon}_{\Lambda} - \Omega_k^2}
\eez
corresponds to the operator $q_t(1,0,0)$ in the renormalized theory.

Combining examples  1-3,  we  find   that   the   expressions   \r{j1}
correspond to correctly defined operators in the renormalized theory.

This work  was supported by the Russian foundation for Basic Research,
project 99-01-01198.

\end{document}